\documentstyle[12pt]{article}
\title{The influence of long-range hopping on ferromagnetism in the
Hubbard model}
\author{Pavol Farka\v sovsk\'y\\
Institute  of  Experimental  Physics,  Slovak   Academy   of
Sciences\\
Watsonova 47, 043 53 Ko\v {s}ice, Slovakia}
\date{}
\begin{document}
\baselineskip=18pt
\maketitle

\begin{abstract}
The phase diagram of the Hubbard model in an external magnetic field
is examined by extrapolation of small-cluster exact-diagonalization
calculations. Using a general expression for the hopping matrix elements
($t_{ij}\sim q^{|i-j|}$) the influence of long-range hopping
(band asymmetry) on ferromagnetism in this model is studied.
It is found that the long-range hopping (nonzero $q$) stabilizes
ferromagnetism in an external magnetic field for $n > 1$.
In the opposite limit $n \leq 1$ the fully polarized ferromagnetic state
is generally suppressed with increasing $q$.
The critical value of magnetic field $h$ below which
the ferromagnetic state becomes unstable is calculated numerically.

\end{abstract}

\newpage

The Hubbard model~\cite{Hubbard} was originally introduced to describe
correlation effects in transition metals, in particular the
band ferromagnetism of Fe, Co and Ni. It soon turned out,
however, that the single-band Hubbard model is not the canonical
model for ferromagnetism. In fact the existence of saturated
ferromagnetism has been proven rigorously only for very special
limits. The first well-known example is the Nagaoka ferromagnetism
that come from the Hubbard model in the limit of infinite repulsion
and one hole in a half-filled band~\cite{Nagaoka}.
Another example where saturated ferromagnetism has been shown to
exist is the case of one dimensional Hubbard model with
nearest and next-nearest neighbor hopping at low electron
densities~\cite{M_H}.
Moreover, several examples of the fully polarized ground state have been
found on special lattices (special conduction bands)
as are the fcc-type lattices~\cite{Ulmke},
the bipartite lattices with sublattices
containing a different number of sites~\cite{Lieb},
the lattices with unconstrained hopping of electrons~\cite{Salerno,Pieri}
and the flat bands~\cite{M_T}.
This indicates that the lattice structure and the kinetic energy of electrons
(the type of hopping) play the crucial role in stabilizing
the ferromagnetic state.  The recent results~\cite{non_bi} obtained on
non-bipartite lattices fully confirm this conjecture.
Non--bipartite lattices have an asymmetric density of states (DOS)
and it is expected that just this asymmetry which brings more
weighage to the one edge of the DOS is the key
to understanding of ferromagnetism in the Hubbard model.

In this paper we want to study how sensible ferromagnetism depends
on the DOS of the non--interacting electrons, particularly
on its asymmetry. To fulfil this program we choose the following
general form for the hopping matrix elements in the one dimension
(the periodic boundary conditions are used)~\cite{Fark}

\begin{equation}
t_{ij}(q)=\left \{ \begin{array}{ll}
  \quad 0,                       &   i=j,\\
  -tq^{|i-j|-1},            &   0 < |i-j| \leq L/2, \\
  -tq^{L-|i-j|-1},          &  L/2 < |i-j \leq L,
\end{array}
\right.
\end{equation}
where $L$ denotes the number of lattice sites and $0\leq q\leq 1$.

Such a selection of hopping matrix elements has several
advantages. It represents a much more realistic
type of electron hopping on a lattice (in comparison to
nearest-neighbor hopping), and it allows us to change
continuously the type of hopping (band) from nearest-neighbor $(q=0)$
to  infinite-range $(q=1)$ hopping and thus immediately
study the effect of the long-range hopping. 
Moreover, for non--zero $q$ the DOS of the non--interacting band 
corresponding to (1) becomes strongly asymmetric since a more weighage 
shifts to the upper edge of the band  with increasing $q$.
Thus one can study simultaneously (by changing only one parameter $q$)
the influence of the increasing  asymmetry in the DOS
and the influence of the long-range hopping on the ground state properties of
the model. Here the special attention is devoted to
the question whether or not the asymmetry in the DOS (the long-range
hopping) can stabilize the ferromagnetic state. As mentioned above the
answer to this question could be the key to understanding of
ferromagnetism in the Hubbard model.

The Hamiltonian of the Hubbard model in an external
magnetic field is given by

\begin{equation}
H=\sum_{ij\sigma}t_{ij}c^+_{i\sigma}c_{j\sigma}+
U\sum_{i}n_{i\uparrow}n_{i\downarrow}-
\frac{h}{2}\sum_{i}(n_{i\uparrow}-n_{i\downarrow}),
\end{equation}
where $c^+_{i\sigma}$ and $c_{i\sigma}$ are the creation and annihilation
operators  for an electron of spin $\sigma$ at site $i$,
$n_{i\sigma}$ is the corresponding number operator
($n_{\sigma}=\frac{1}{L}\sum_i n_{i\sigma}$),
$U$ is the on-site Coulomb interaction constant and
$h$ is an external magnetic field.

The first rigorous criteria for the stability of ferromagnetism
in this model (for $q=0$ and $n=n_{\uparrow}+n_{\downarrow}$=1)
have been derived by Strack and Vollhardt~\cite{Strack}.
Their results provide the rigorous upper
bound on the critical magnetic field above which the fully
polarized ferromagnet is the exact ground state. The lower bound
on the magnetic field below which the ferromagnetic state becomes
locally unstable has been calculated by van Dongen and
Jani\v{s}~\cite{van}. Very recently some exact results about
the ground state of the Hubbard model with an infinite-range
hopping ($q=1$) appeared in literature~\cite{Salerno,Pieri}.
For the infinite-range hopping and just one electron more
than half-filling Pieri~\cite{Pieri} proved rigorously that the ground
state of the model is the Nagaoka ferromagnetic state for every
value of $U>0$. The limit of infinite-range hopping is, however,
the least realistic limit of Eq.~1. It is  interesting, therefore,
to look on the possibility of ferromagnetism in the Hubbard model
with a generalized type of hopping for smaller values of $q$
that describe much more realistic type of electron hopping.

In this paper we extend the calculations to arbitrary $q$
and arbitrary band filling~$n$.
The ground state properties  of the model, and particularly
ferromagnetism are studied numerically (using the Lanczos method)
for a wide range of parameters ($q,U,h,n$)
and typical examples are then chosen from a large number of
available results to represent the most interesting cases.
The results obtained are presented in the form of phase
diagrams in the $h$-$U$ plane. To determine the phase diagram in the $h$-$U$
plane (corresponding to some $q$ and $n$) the up-spin (down-spin) electron
occupation number $n_\uparrow$ ($n_\downarrow$) are calculated
point by point as functions of $h$ and $U$.
Of course, such a procedure demand
in practice a considerable amount of CPU time which
impose severe restrictions on the size of
clusters that can be studied with this method~($L \sim 14$).
However, we will show later that even the study of such small
clusters can reveal some general features of the model.

First we have investigated the model for  small finite clusters
(up to 12 sites) at half-filling.
The most important  result  obtained  for  the
following  set  of $q$ values: $q=0,0.2,0.4,0.6,0.8$ is that
the fully polarized ferromagnetic state  ($n_{\uparrow}=1,n_{\downarrow}=0$)
is suppressed with increasing $q$.
Numerical results for the critical magnetic field $h_{c}(U)$ above which
the ground state is fully polarized ferromagnet are shown in Fig.~1.
To determine the finite size effect the critical behavior $h_{c}(U)$
has been computed for several values of $L$ ($L=6,8,10$ and $12$),
but no significant finite-size effects have been observed
over the whole interval of $U$ plotted. Therefore we
suppose that these results can be satisfactory extended to large
systems. This conjecture supports also the analytical calculation
of $h_c$ from the well-known single spin-flip ansatz.

The single spin-flip ansatz for the Hamiltonian~(2) is~\cite{van}

\begin{equation}
\Psi=\sum_{n,m}a_{nm}c^+_{n\downarrow}c_{m\uparrow}|F\rangle,
\end{equation}
where $|F\rangle=\prod_i c^+_{i\uparrow}|0\rangle$ is the fully
polarized ferromagnetic state. Inserting $|\Psi\rangle$ into the
Schr\"odinger equation yields the consistence relation
\begin{equation}
1=\frac{U}{L}\sum_{\bf k} \frac{1}{\epsilon_{\bf k}
-\epsilon_{{\bf K}-{\bf k}}-E+h+U}
\end{equation}
from which the critical magnetic field $h_c$ can be determined
immediately. Since the dispersion relation
$\epsilon_{\bf k}=\frac{1}{L}\sum_{ij}t_{ij}e^
{-i{\bf k}({\bf R}_i-{\bf R}_j)}$ for the generalized
type of electron hopping~(1) is much more complicated than one
corresponding to nearest-neighbor hopping we have solved~(4) numerically.
The results of these numerical calculations (ssf) are compared in
Fig.~1 with small-cluster exact-diagonalization calculations.
The accordance of results is very good, particularly for $q \leq 0.5$,
indicating that (i) finite-size effects are small, and (ii) the single
spin-flip ansatz is a good approach for the half-filled band case.
We have observed the same behavior of the model also for $n<1$
which leads to the conclusion that the long-range hopping of the
type~(1) does not support ferromagnetism in the Hubbard model at
least in the region $n \leq 1$.

It should be noted that our findings are fully consistent with results
of Salerno~\cite{Salerno} and Pieri~\cite{Pieri}. They found no ferromagnetic
ground state in the Hubbard model with infinite-range hopping $(q=1)$ for
$n\leq 1$ and $h=0$. Therefore we have turned our attention to the case $n>1$.

The results of small-cluster exact-diagonalization calculations
obtained for $n=3/2$ and different values of $q$ are shown in
Fig.~2 and Fig.~3. To reveal the finite-size effects we have displayed
results for two finite clusters of $L=8$ and $L=14$ sites.
It is seen that the finite-size effects are now non-zero,
but  in spite of this fact some general features
of the model are obvious. The most important result is that
the ferromagnetic state is stabilized with increasing
asymmetry of the DOS. The influence of asymmetry is so strong that for
$q$ sufficiently large the transition to the ferromagnetic state
can be induced even at $h=0$.
This result confirms the conjecture that the asymmetry of the DOS
(the long-range hopping) plays the crucial role in stabilizing the
ferromagnetic state at $h=0$, at least on finite clusters.
Unfortunately, the size of clusters that can be studied with the Lanczos
method is too small to show satisfactory whether or not this important
result persists also in the thermodynamic limit $L \to \infty$.
To resolve this problem some other methods that allow to work with
larger clusters (e.g. the density-matrix renormalization group)
should be used.
Work in this direction in currently in progress.

Another important result that can be found using small-cluster
exact-diago\-na\-li\-za\-tion calculations is shown in Fig.~3. There are
compared small-cluster exact-diago\-na\-liza\-tion results with numerical
results obtained from the single spin-flip ansatz. It is seen that the
results strongly differ, particularly for $U$ sufficiently large.
This indicates that the transition to the ferromagnetic state is
probably discontinuous at least in the strong coupling limit.
At the same time this result explains why the critical
interaction strength $U_c(h=0)$ obtained from the single spin-flip
ansatz lies usually much lower than the exact U$_c$~\cite{Dual}.
For $n>1$ the single spin-flip ansatz is not a good approach to
determine the ferromagnetic region.

In summary, the numerical study of the Hubbard model in an external
magnetic field shows that the long-range hopping (band asymmetry)
stabilizes ferromagnetism in this model for $n>1$. In the opposite
limit $n \leq 1$ the fully polarized ferromagnetic state is generally
suppressed with increasing band asymmetry. In addition, small-cluster
exact-diagonalization calculations indicate that the transition to
the ferromagnetic state (for $n>1$) is probably discontinuous at
least in the strong coupling limit.
\vspace{0.5cm}

This work was partially supported by the Slovak Grant Agency for
Science and the Commission of the European Communities Contract
No. CIPA-CT93-0183.

\newpage
Figure Captions

\vspace{0.5cm}
Fig. 1. The critical magnetic field $h_{c}$ as a function of $U$
calculated from the single spin-flip ansatz (ssf) and exactly (exact)
for $n=1$.
Curves (from down to up) correspond to $q=0, q=0.2, q=0.4,$ and $q=0.6$.
For $L=10$ ssf and exact-diagonalization results are identical.
\vspace{0.5cm}

Fig. 2. The critical magnetic field $h_{c}$ as a function of $U$
calculated exactly for $L=8$ and $n=3/2$.
Curves (from up to down) correspond to $q=0, q=0.2,$ $q=0.4, q=0.6$ and $q=0.8$.
\vspace{0.5cm}

Fig. 3. The critical magnetic field $h_{c}$ as a function of $U$
calculated from the single spin-flip ansatz (ssf) and exactly (exact)
for $L=14$ and $n=3/2$.
Curves (from up to down) correspond to $q=0, q=0.2, q=0.4,$ and $q=0.6$.

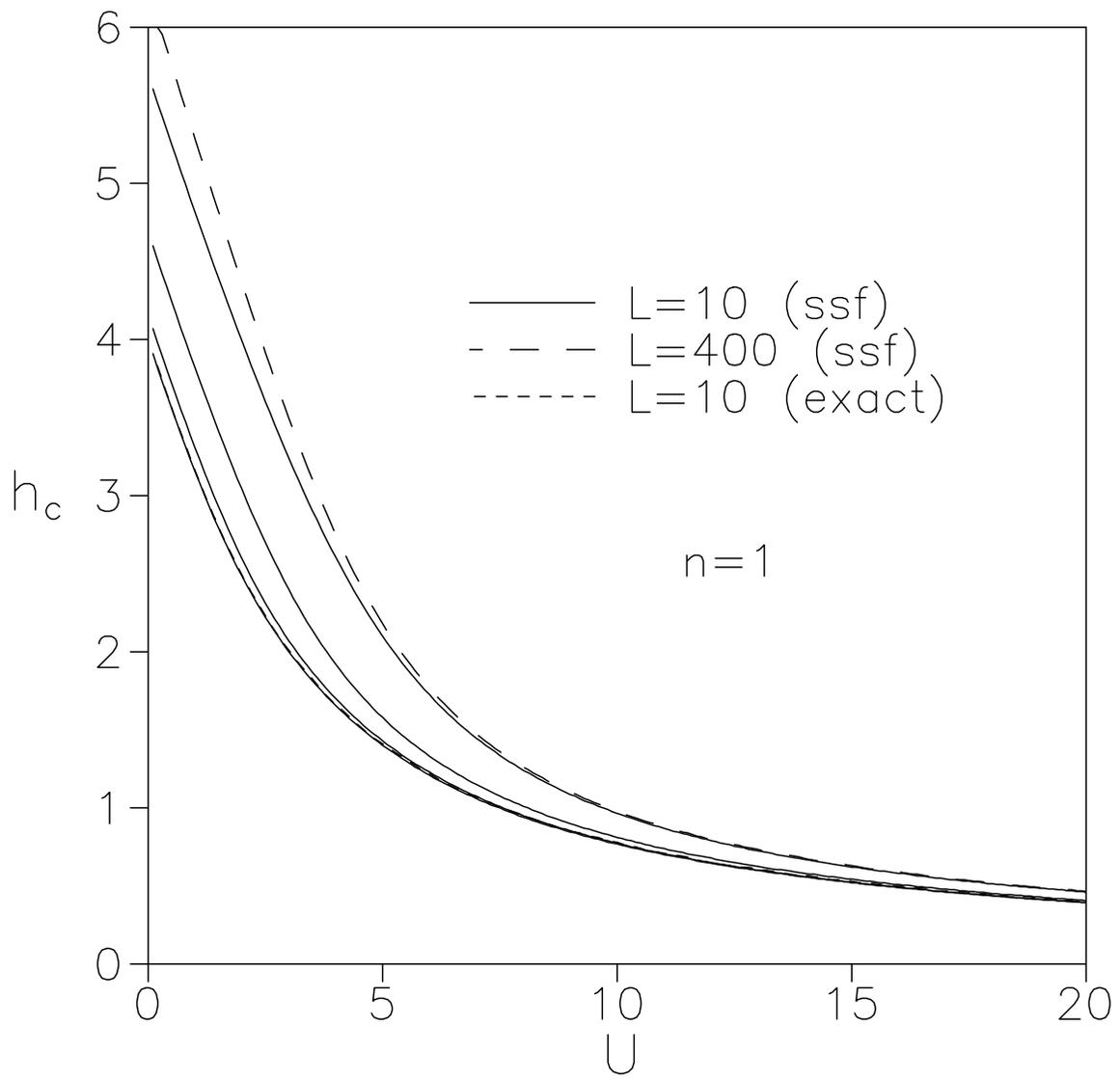
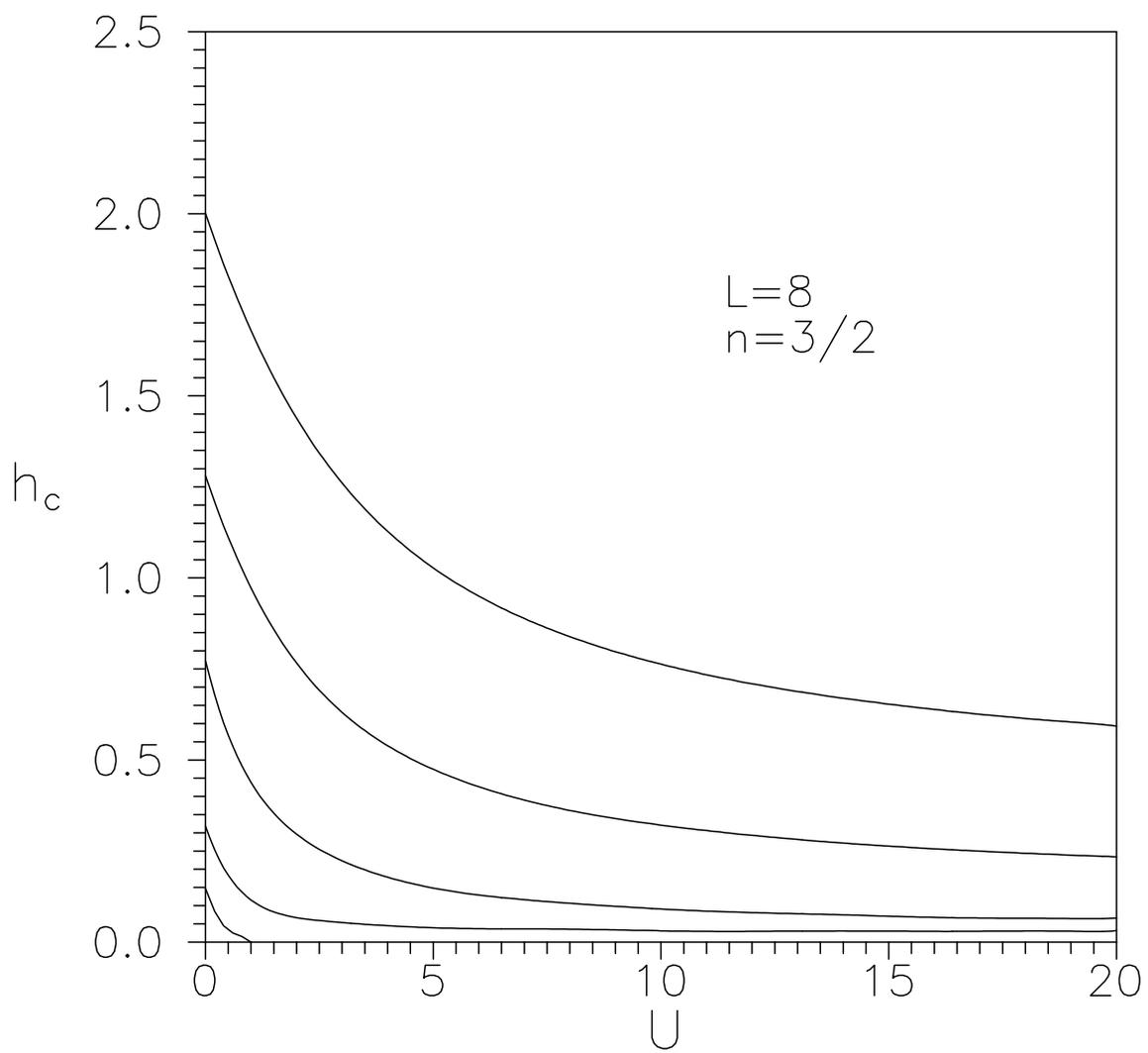
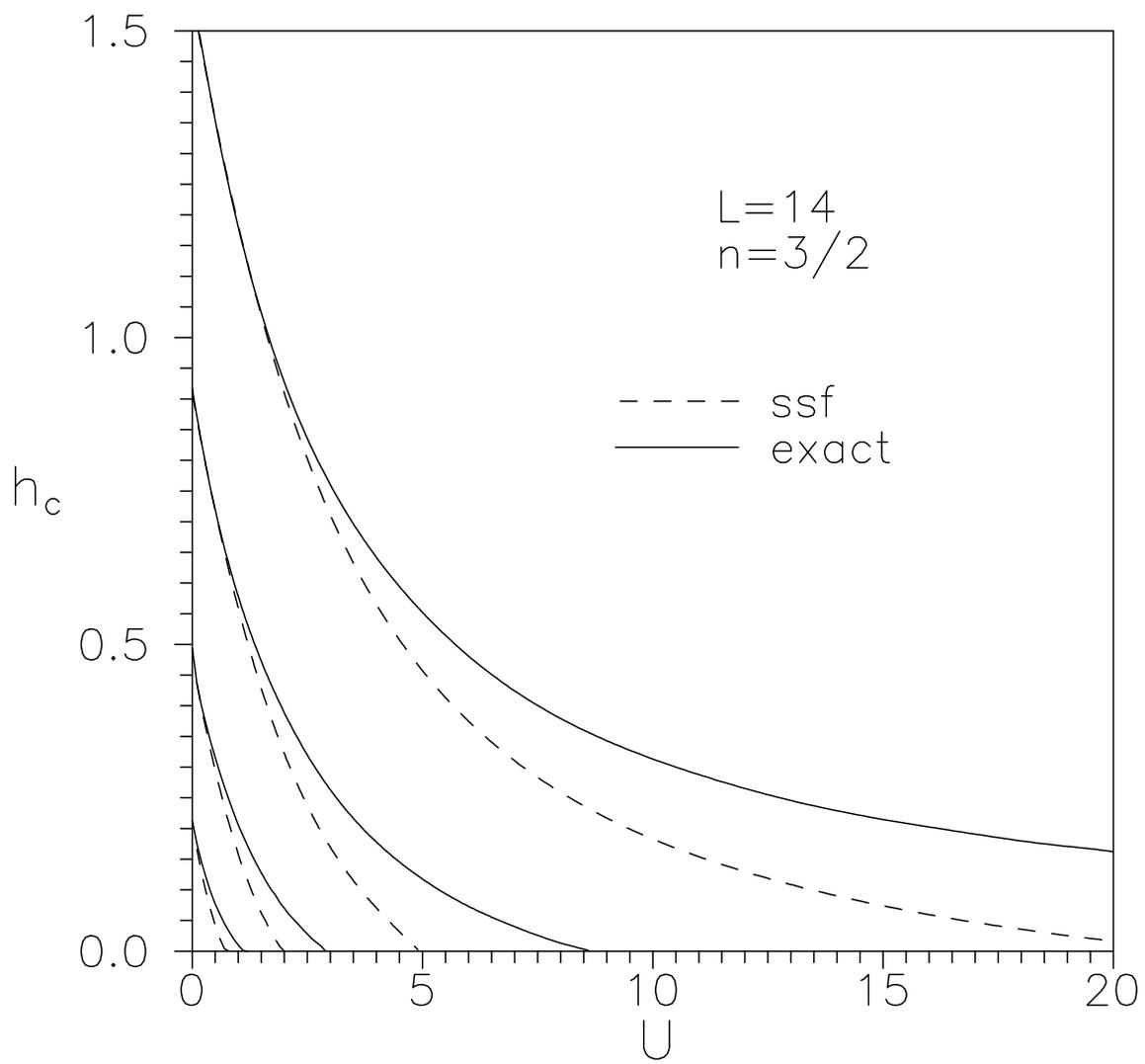
\end{document}